\newfam\msbfam
\font\twlmsb=msbm10 at 12pt
\font\eightmsb=msbm10 at 8pt
\font\sixmsb=msbm10 at 6pt
\textfont\msbfam=\twlmsb
\scriptfont\msbfam=\eightmsb
\scriptscriptfont\msbfam=\sixmsb
\def\cj{\fam\msbfam}

\def\C{{\cj C}}

\def\Z{{\cj Z}}

\centerline{\bf REMARK ON CHARGE CONJUGATION IN THE NON RELATIVISTIC LIMIT}

\

\centerline{\bf A. Cabo$^1$, D. B. Cervantes$^2$, H. P\'erez Rojas$^1$ and M. Socolovsky$^2$} 

\

\centerline{\it $^1$ Grupo de F\'\i sica Te\'orica, ICIMAF, Calle E, Nr. 309, Vedado, Habana 4, Cuba}

\centerline{\it $^2$ Instituto de Ciencias Nucleares, Universidad Nacional Aut\'onoma de M\'exico}
\centerline{\it Circuito Exterior, Ciudad Universitaria, 04510, M\'exico D. F., M\'exico} 

\

{\it We study the non relativistic limit of the charge conjugation operation $\cal C$ in the context of the Dirac equation coupled to an electromagnetic field. The limit is well defined and, as in the relativistic case, $\cal C$, $\cal P$ (parity) and $\cal T$ (time reversal) are the generators of a matrix group isomorphic to a semidirect sum of the dihedral group of eight elements and $\Z_2$. The existence of the limit is supported by an argument based in  quantum field theory. Also, and most important, the limit exists in the context of galilean relativity. Finally, if one complexifies the Lorentz group and therefore the galilean spacetime $x_\mu$, then the explicit form of the matrix for $\cal C$ allows to interpret it, in this context, as the complex conjugation of the spatial coordinates: $\vec{x} \to \vec{x}^*$. This result is natural in a fiber bundle description.}

\


\

1. {\bf Introduction}

\

It is generally believed that the concept of antiparticles can only be defined in the context of relativistic quantum mechanics (RQM). The basic reason is that it is only in this regime that one can have free particles with negative energies travelling backwards in time, whose absence is interpreted as positive energy and opposite charge and momentum particles travelling forwards in time: {\it antiparticles}. Then, in particular, the operation of {\it charge conjugation} $\cal C$ which makes the transformation particle $\longleftrightarrow$ antiparticle should only exist in RQM $^{1,2}$, and therefore in the context of the Poincar\'e group. It is well known, however, that $\cal C$ does {\it not} belong to this group. 

One can, however, give an {\it ad hoc} definition of $\cal C$, not only in the context of non relativistic quantum mechanics but also in classical lorentzian and galilean mechanics $^{3}$. Nevertheless, for consistency, any prescription in the non relativistic approximation should be derived from the relativistic theory as the limiting case $\vert \beta \vert <<1$ where $\beta$ is the particle "velocity" ($c$=1). 

In this note and in the concrete case of a Dirac field coupled to an external electromagnetic field, we prove that the above mentioned {\it ad hoc} prescription for $\cal C$ can indeed be obtained from first principles; that is, the charge conjugation symmetry of the relativistic wave equation leads, in the non relativistic approximation, to fermionic wave functions respectively describing low energy electrons and positrons, related to each other by the matrix representing $\cal C$ in this approximation, $C_{nr}$. The most important contribution of this note, however, is that in the non relativistic limit, we can show that $\cal C$ {\it exists in the context of the galilean group of transformations} (galilean relativity).  In section 3, we explicitly prove the galilean invariance of the Schroedinger-Pauli equations for Dirac particles and antiparticles in this limit, and that these equations are related to each other by the limiting matrix $C_{nr}$. 

Qualitatively this occurs because, as mentioned above, in the quantum relativistic theory there is a symmetry between the particle and antiparticle worlds. Both, the particle equation and the antiparticle equation, have well defined non relativistic limits and one should not expect a contradiction between them, for example that the field components should vanish as $\vert \beta \vert \to 0$. It is quite obvious that if one has e.g. low energy electrons, one should also have low energy positrons, with electrons and positrons related by $\cal C$. This result, therefore, invalidates the usual claim that $\cal C$ can only be considered a relativistic symmetry.

\

2. {\bf Charge conjugation}

\

The wave equation for a spinor field $$\psi=\pmatrix{\psi_1 \cr \psi_2 \cr \psi_3 \cr \psi_4 \cr}\equiv \pmatrix{\varphi \cr \chi \cr} \ \ with \ \ \varphi=\pmatrix{\psi_1 \cr \psi_2 \cr} \ \ and \ \ \chi=\pmatrix{\psi_3 \cr \psi_4 \cr}$$ of electric charge $q$ and mass $m$, coupled to an external electromagnetic 4-potential $A^\mu=(V,\vec{A})$, is given by $$(i\gamma^\mu\partial_\mu-qA_\mu\gamma^\mu-m)\psi=0 \eqno{(1)}$$ where $\gamma^\mu$, $\mu=0,1,2,3$, are the Dirac matrices. In terms of the matrices $\vec{\alpha}=\gamma^0\vec{\gamma}$ and in the standard (Pauli-Dirac) representation $^{4}$, equation (1) is $$i\partial_t\psi=(\vec{\alpha}\cdot(\vec{p}-q\vec{A})+q\phi+m\gamma_0)\psi \eqno{(2)}$$ where $\vec{p}=-i\nabla$, $\vec{\alpha}=\pmatrix{0 & \vec{\sigma} \cr \vec{\sigma} & 0 \cr}$, $\gamma_0=\pmatrix{1 & 0 \cr 0 & -1 \cr}$, $A_\mu=(\phi,-\vec{A})$, and $\vec{\sigma}$ are the Pauli matrices. The charge conjugate spinor $\psi_C$ is given by $$\psi_C=C {\bar{\psi}}^\sim =\pmatrix{\psi^*_4 \cr -\psi^*_3 \cr -\psi^*_2 \cr \psi^*_1 \cr}\equiv \pmatrix{\varphi_C \cr \chi_C \cr} \eqno{(3)}$$ where $\varphi_C=\pmatrix{\psi^*_4 \cr -\psi^*_3 \cr}$ and $\chi_C=\pmatrix{-\psi^*_2 \cr \psi^*_1 \cr}$, $\bar{\psi}=\psi^\dagger \gamma_0$ is the Dirac conjugate spinor, $C$ is the charge conjugation matrix which up to a sign is given by $^{5}$ $$C=i\gamma^2\gamma_0=\pmatrix{0 & 0 & 0 & -1 \cr 0 & 0 & 1 & 0 \cr 0 & -1 & 0 & 0 \cr 1 & 0 & 0 & 0 \cr}, \eqno{(4)}$$ and $\sim$ denotes the transpose vector. $\psi_C$ obeys the equation $$(i\gamma^\mu\partial_\mu+qA_\mu \gamma^\mu-m)\psi_C=0, \eqno{(5)}$$ the analogous of (2) being $$i\partial_t\psi_C=(\vec{\alpha}\cdot (\vec{p}+q\vec{A})-q\phi+m\gamma_0)\psi_C. \eqno{(6)}$$ Clearly, $\psi_C$ describes particles with the same mass but opposite charge. If the charge conjugation transformation is completed with the replacement $$A_\mu \buildrel \ {\cal C} \over \longrightarrow -A_\mu \eqno{(7)}$$ then (5) and (6) have the same form as (1) and (2) exhibiting the complete symmetry of quantum electrodynamics under $\cal C$.

\

3. {\bf Non relativistic limit}

\

Defining $\tilde{\psi}$ through $$\psi(\vec{x},t)=e^{-imt}\tilde{\psi}(\vec{x},t)\equiv e^{-imt}\pmatrix{\tilde{\varphi} \cr \tilde{\chi} \cr}, \eqno{(8)}$$ where the exponential factor involves the {\it positive} rest energy $m$ ($mc^2$), equation (2) is equivalent to the system of equations $$i\partial_t\tilde{\varphi}=\vec{\sigma}\cdot\vec{\pi}\tilde{\chi}+q\phi\tilde{\varphi}, \eqno{(9a)} $$ $$i\partial_t\tilde{\chi}=\vec{\sigma}\cdot \vec{\pi}\tilde{\varphi}+q\phi\tilde{\chi}-2m\tilde{\chi} \eqno{(9b)}$$ with $\vec{\pi}=\vec{p}-q\vec{A}$. {\it In the non relativistic approximation}, $m$ is the largest energy $^{4}$, and one can neglect the terms $i\partial_t\tilde{\chi}$ and $qV\tilde{\chi}$ in equation (9b), obtaining the "small" components $\tilde{\chi}$ in terms of the large components $\tilde{\varphi}$: $$\tilde{\chi}={{\vec{\sigma}\cdot \vec{\pi}}\over {2m}}\tilde{\varphi}. \eqno{(10)}$$ Clearly, $\vert \vert \tilde{\chi} \vert \vert /\vert \vert \tilde{\varphi} \vert \vert \simeq {{\vert \vec{\pi}\vert}\over {m}}\simeq \beta <<1$. Replacing (10) in (9a) leads to the Schroedinger-Pauli equation for the two-component spinor $\varphi$: $$i{{\partial}\over{\partial t}}\pmatrix{\tilde{\psi}_1 \cr \tilde{\psi}_2 \cr}={{1}\over{2m}}(-\nabla^2+{{q^2}\over{c^2}}\vec{A}^2+{{iq}\over{2mc}}\nabla\cdot\vec{A}+2i{{q}\over{c}}\vec{A}\cdot\nabla-{{q}\over{c}}\vec{\sigma}\cdot \vec{B}+2mq\phi)\pmatrix{\tilde{\psi}_1\cr\tilde{\psi}_2 \cr}, \eqno{(11)}$$ where $\vec{B}=\nabla \times \vec{A}$, and we have restored $c$, the velocity of light in vacuum.  
For the absolute values of the spinor components $\psi_a$, $a=1,2,3,4$ one obtains $$\vert \tilde{\psi}_3 \vert, \ \vert \tilde{\psi}_4 \vert <<\vert \tilde{\psi}_1 \vert, \ \vert \tilde{\psi}_2\vert. \eqno{(12)}$$ If one restores $c$, this means that $$\vert \tilde{\psi}_3 \vert, \ \vert \tilde{\psi}_4 \vert \to 0 \ as \ c \to \infty. \eqno{(13)}$$

\

The corresponding equations for the charge conjugate spinor $\psi_C$, $$\psi_C=e^{imt}\tilde{\psi}_C=e^{imt}\pmatrix{\tilde{\varphi}_C \cr \tilde{\chi}_C \cr}, \eqno{(14)}$$ where now the exponential factor involves the {\it negative} rest energy $-m$ ($-mc^2$), are $$i\partial_t \tilde{\varphi}_C=\vec{\sigma}\cdot \vec{\pi}^\prime \tilde{\chi}_C+(-q\phi+2m)\tilde{\varphi}_C, \eqno{(15a)}$$ $$i\partial_t\tilde{\chi}_C=\vec{\sigma}\cdot\vec{\pi}^\prime \tilde{\varphi}_C-q\phi\tilde{\chi}_C, \eqno{(15b)}$$ with $\vec{\pi}^\prime=\vec{p}+q\vec{A}$. Again, in the non relativistic approximation, $$\tilde{\varphi}_C=-{{\vec{\sigma}\cdot \vec{\pi}^\prime}\over {2m}}\tilde{\chi}_C \eqno{(16)}$$ with $$\tilde{\varphi}_C=\pmatrix{\tilde{\psi}^*_4 \cr -\tilde{\psi}^*_3 \cr} \ , \ \tilde{\chi}_C=\pmatrix{-\tilde{\psi}^*_2 \cr \tilde{\psi}^*_1 \cr} \eqno{(17)}$$ and therefore $$\vert \vert \tilde{\varphi}_C \vert \vert /\vert \vert \tilde{\chi}_C\vert \vert \simeq \beta<<1. \eqno{(18)}$$ Clearly, replacing (16) in (15b) leads to the Schroedinger-Pauli equation for $\tilde{\chi}_C$: $$i{{\partial}\over{\partial t}}\pmatrix{-\tilde{\psi}_2^*\cr \tilde{\psi}_1^* \cr}={{1}\over{2m}}(\nabla^2-{{q^2}\over{c^2}}\vec{A}^2+{{iq}\over{2mc}}\nabla\cdot\vec{A}+2i{{q}\over{c}}\vec{A}\cdot\nabla-{{q}\over{c}}\vec{\sigma}\cdot\vec{B}-2mq\phi)\pmatrix{-\tilde{\psi}_2^* \cr \tilde{\psi}_1^* \cr}. \eqno{(19)}$$

\

Also, and most important, (18) is consistent with (12); this allows a natural definition of the charge conjugation matrix (operator in the 2-dimensional Hilbert space $\C^2$) in the non relativistic limit:$$\pmatrix{\tilde{\psi}_1 \cr \tilde{\psi}_2 \cr}\equiv \pmatrix{u \cr v \cr} \buildrel {\cal C} \over \longrightarrow \pmatrix{-\tilde{\psi}^*_2 \cr \tilde{\psi}^*_1 \cr}\equiv \pmatrix{-v^* \cr u^* \cr}:=C_{nr}\pmatrix{u \cr v \cr} \eqno{(20)}$$ with $$C_{nr}=K\pmatrix{0 & -1 \cr 1 & 0 \cr} \eqno{(20a)}$$ where $K$ is the complex conjugation operation. Since $C_{nr}(\lambda\pmatrix{u \cr v \cr})=\lambda^*C_{nr}\pmatrix{u \cr v \cr}$, then $C_{nr}$ is antilinear. $C_{nr}$ has the following properties: $$C_{nr}^2=-1, \ C_{nr}^{-1}=-C_{nr}=C_{nr}^\sim=-C_{nr}^*=C_{nr}^\dagger, \eqno{(21)}$$ which are the same properties of the relativistic matrix. Together with the matrices $\hat{P}\equiv P_{nr}=i\pmatrix{1 & 0 \cr 0 & 1 \cr}$ (parity) and $\hat{T}\equiv T_{nr}=\pmatrix{0 & -1 \cr 1 & 0 \cr}$ (time reversal) of reference 6, $C_{nr}$, $P_{nr}$ and $T_{nr}$ have the multiplication table $$\matrix{ & C_{nr} & P_{nr} & T_{nr} \cr C_{nr} & -1 & C_{nr}P_{nr} & C_{nr}T_{nr} \cr P_{nr} & -C_{nr}P_{nr} & -1 & P_{nr}T_{nr} \cr T_{nr} & C_{nr}T_{nr} & P_{nr}T_{nr} & -1 \cr}$$ from which, using associativity, one obtains the same group multiplication table as in the relativistic case$^5$, namely, the group $16E$, which is isomorphic to a semidirect sum of $DH_8$, the dihedral group of eight elements, and $\Z_2$, the 0-sphere. 

\

It is important to emphasize that the definition of $C_{nr}$ is indeed in the context of {\it galilean relativity}. In fact, to prove it, under the galilean transformations of inertial reference systems $S(t, x,y,z)$ and $S^\prime (t^\prime,x\prime,y^\prime, z^\prime)$ with $t^\prime=t$ and $\vec{x}=\vec{x}^\prime+\vec{V}t^\prime$, which imply ${{\partial}\over{\partial t}}={{\partial}\over{\partial t^\prime}}-\vec{V}\cdot{{\partial}\over{\partial \vec{x}^\prime}}$ and $\nabla=\nabla^\prime$ (with $\nabla={{\partial}\over{\partial\vec{x}}}$ and $\nabla^\prime={{\partial}\over{\partial\vec{x}^\prime}}$), the change of the wave functions $\pmatrix{u(\vec{x},t) \cr v(\vec{x},t)\cr}$ and $\pmatrix{-v^*(\vec{x},t)\cr u^*(\vec{x},t)}$ are respectively given by$^7$ $$\pmatrix{u(\vec{x},t) \cr v(\vec{x},t)\cr}=e^{im(\vec{V}\cdot\vec{x}^\prime+{{1}\over{2}}\vert\vec{V}\vert ^2 t^\prime)}\pmatrix{\tilde{u}(\vec{x}^\prime,t^\prime) \cr \tilde{v}(\vec{x}^\prime,t^\prime)\cr} \eqno{(22a)}$$ and $$\pmatrix{-v^*(\vec{x},t) \cr u^*(\vec{x},t)}=e^{-im(\vec{V}\cdot\vec{x}^\prime+{{1}\over{2}}\vert \vec{V}\vert ^2 t^\prime)}\pmatrix{-\tilde{v}^*(\vec{x}^\prime,t^\prime) \cr \tilde{u}^*(\vec{x}^\prime,t^\prime)\cr}. \eqno{(22b)}$$ Restricting for simplicity to boosts in direction $x$, {\it i.e.} $\vec{V}=V^1 \hat{e}_1$, and using the relativistic transformation of the 4-potential $A^\mu$ and the associated electric and magnetic field strengths $\vec{E}$ and $\vec{B}$ to order $\beta$ ({\it i.e.} neglecting terms of order $\beta^2$ and higher), $$(\phi,A^1,A^2,A^3)=(\phi^\prime+\beta {A^1}^\prime,{A^1}^\prime+\beta\phi^\prime,{A^2}^\prime,{A^3}^\prime), \eqno{(23a)}$$ $$(E^1,E^2,E^3)=({E^1}^\prime, {E^2}^\prime-\beta {E^3}^\prime, {E^3}^\prime+\beta{B^2}^\prime), \eqno{(23b)}$$ $$(B^1,B^2,B^3)=({B^1}^\prime,{B^2}^\prime+\beta {E^3}^\prime, {B^3}^\prime-\beta{E^2}^\prime), \eqno{(23c)}$$ the equations (11) and (19) transform into (we reinsert the dependence on $c$) $$i{{\partial}\over{\partial t^\prime}}\pmatrix{\tilde{u} \cr \tilde{v}\cr}=(-{{{\nabla^\prime}^2}\over{2m}}+{{q^2}\over{2mc^2}}{\vec{A^\prime}}^2+{{iq}\over{mc}}\vec{A}^\prime\cdot\nabla^\prime+{{iq}\over{2mc}}\nabla^\prime\cdot\vec{A}^\prime-{{q}\over{2mc}}\vec{\sigma}\cdot\vec{B}^\prime+q\phi^\prime)\pmatrix{\tilde{u}\cr \tilde{v}\cr}$$ $$+\beta({{q^2}\over{mc^2}}{A^1}^\prime\phi^\prime+{{iq}\over{mc}}\phi^\prime {{\partial}\over{\partial {x^1}^\prime}}-{{q}\over{c}}\phi^\prime V^1+{{iq}\over{2mc}}{{\partial}\over{\partial {x^1}^\prime}}\phi^\prime-{{q}\over{2mc}}(\vec{\sigma}\times\vec{E}^\prime)_1)\pmatrix{\tilde{u}\cr \tilde{v}\cr} \eqno{(24a)}$$ and $$i{{\partial}\over{\partial t^\prime}}\pmatrix{-\tilde{v}^* \cr \tilde{u}^*\cr}=({{{\nabla^\prime}^2}\over{2m}}-{{q^2}\over{2mc^2}}{\vec{A^\prime}}^2+{{iq}\over{mc}}\vec{A}^\prime \cdot\nabla^\prime+{{iq}\over{2mc}}\nabla^\prime\cdot\vec{A}^\prime-{{q}\over{2mc}}\vec{\sigma}\cdot\vec{B}^\prime-q\phi^\prime)\pmatrix{-\tilde{v}^*\cr \tilde{u}^*\cr}$$ $$+\beta(-{{q^2}\over{mc^2}}{A^1}^\prime\phi^\prime+{{iq}\over{mc}}\phi^\prime{{\partial}\over{\partial {x^1}\prime}}+{{q}\over{c}}\phi^\prime V^1+{{iq}\over{2mc}}{{\partial}\over{\partial {x^1}^\prime}}\phi^\prime-{{q}\over{2mc}}(\vec{\sigma}\times\vec{E}^\prime)_1)\pmatrix{-\tilde{v}^*\cr \tilde{u}^*\cr}. \eqno{(24b)}$$ To be consistent with the approximation leading from equations (9a,b) and (15a,b) to (11) and (19), respectively, the terms proportional to $\beta$ in the right hand sides of equations (24a,b) have to be neglected, thus exhibiting the {\it galilean invariance of the Schroedinger-Pauli approximation to the Dirac equation}. 

\

Finally, it is crucial to remark that the non relativistic equations (11) and (19) are transformed into each other by the application of the matrix operator $C_{nr}$, in the same way that $C$ transforms (1) into (5) in the relativistic case, thus reflecting the {\it galilean character of the approximation $C_{nr}$ to $C$}. (To prove this, we multiply equation (11) from the left by $C_{nr}$, insert the unity $-C_{nr}^2$ between the factors $-{{q}\over{c}}\vec{\sigma}\cdot \vec{B}$ and $\pmatrix{\tilde{\psi}_1 \cr \tilde{\psi}_2 \cr}$, and use the fact that $K\sigma_l K=\sigma_l$ for $l=1,3$ and $K\sigma_2 K=\sigma_2^*$.)

\

4. {\bf Field theory argument}

\

To see the possibility of defining the quantum operator $\bf C$ corresponding to $\cal C$ in the non relativistic approximation, it is enough to study the finite part of the energy density operator of the free Dirac field: $$p_0=\int {{d^3\vec{k}}\over{(2\pi)^3}}k_0\sum_{\alpha=1}^2(N_\alpha(\vec{k})+\bar{N}_\alpha(\vec{k})) \eqno{(25)}$$ where $k_0=\sqrt{\vec{k}^2+m^2}$ and $N_\alpha(\vec{k})$ and $\bar{N}_\alpha(\vec{k})$ are respectively the occupation number operators for electrons and positrons, related to the corresponding creation and annihilation operators through $$N_\alpha(\vec{k})(2\pi)^3\delta^3(\vec{0}){{k_0}\over{m}}=b^\dagger_\alpha(\vec{k})b_\alpha(\vec{k}), \eqno{(26a)}$$ $$\bar{N}_\alpha(\vec{k})(2\pi)^3\delta^3(\vec{0}){{k_0}\over{m}}=d^\dagger_\alpha(\vec{k})d_\alpha(\vec{k}) \eqno{(26b)}$$ in the infinite volume limit $V=(2\pi)^3\delta^3(\vec{0})=\infty$.

\

In the non relativistic approximation, $$k_0\simeq m+{{\vert \vec{k} \vert ^2} \over{2m}} \eqno{(27)}$$ and, up to an infinite constant operator corresponding to the rest energy, the non relativistic energy density operator for the electron-positron system is given by $$p_0^{nr}=\int {{d^3\vec{k}}\over{(2\pi)^3}}{{\vert\vec{k}\vert^2}\over{2m}}\sum_{\alpha=1}^2(N_\alpha(\vec{k})+\bar{N}_\alpha(\vec{k})). \eqno{(28)}$$ It is easy to verify that this hamiltonian density is invariant under the operator $^{8}$ $${\bf C}=\Pi_{\vec{k},\alpha}{\bf C}_{\vec{k},\alpha}, \eqno{(29)}$$ where $${\bf C}_{\vec{k},\alpha}=1-2\beta^\dagger_{\vec{k},\alpha}\beta_{\vec{k},\alpha} \eqno{(30)}$$ and $$\beta_{\vec{k},\alpha}={{1}\over{\sqrt{2}}}(b_{\vec{k},\alpha}-d_{\vec{k},\alpha}). \eqno{(31a)}$$ Clearly, $${\bf C}_{\vec{k},\alpha}{\bf C}_{\vec{k}^\prime,\alpha^\prime}={\bf C}_{\vec{k}^\prime,\alpha^\prime}{\bf C}_{\vec{k},\alpha} \eqno{(32)}$$ if $(\vec{k},\alpha)\neq(\vec{k}^\prime,\alpha^\prime)$. Then $${\bf C}^2=1. \eqno{(33)}$$ Also, $${\bf C}^\dagger={\bf C}^{-1}={\bf C} \eqno{(34)}$$ {\it i.e.} ${\bf C}$ is unitary and hermitian. Since $${\bf C}\alpha_{\vec{k},\alpha}{\bf C}^{-1}=\alpha_{\vec{k},\alpha} \ \ and \ \ {\bf C}\beta_{\vec{k},\alpha}{\bf C}^{-1}=-\beta_{\vec{k},\alpha} \eqno{(35)}$$ where $$\alpha_{\vec{k},\alpha}={{1}\over {\sqrt{2}}}(b_{\vec{k},\alpha}+d_{\vec{k},\alpha}), \eqno{(31b)}$$ then $${\bf C}b_{\vec{k},\alpha}{\bf C}^{-1}=d_{\vec{k},\alpha}, \ \ {\bf C}d_{\vec{k},\alpha}{\bf C}^{-1}=b_{\vec{k},\alpha}, \eqno{(36a)}$$ $${\bf C}b^\dagger_{\vec{k},\alpha}{\bf C}^{-1}=d^\dagger_{\vec{k},\alpha}, \ \ {\bf C}d^\dagger_{\vec{k},\alpha}{\bf C}^{-1}=b^\dagger_{\vec{k},\alpha}, \eqno{(36b)}$$ that is, ${\bf C}$ changes particles into antiparticles and viceversa, and therefore $${\bf C}N_{\vec{k},\alpha}{\bf C}^{-1}=\bar{N}_{\vec{k},\alpha}, \ \ {\bf C}\bar{N}_{\vec{k},\alpha}{\bf C}^{-1}=N_{\vec{k},\alpha} \eqno{(37)}$$ leaving $p_0^{nr}$ invariant. So, {\it also in the non relativistic limit}, ${\bf C}$ {\it can be identified with the charge conjugation operator}.

This result is also supported by the fact that the (electric) charge operator $\hat{q}$ is also well defined in the non relativistic limit. In fact, its finite part for the free Dirac field with charge $e$ ($e<0$) is given by $$\hat{q}_0=e\int{{d^3\vec{k}}\over {(2\pi)^3}}({{m}\over{k_0}})\sum^2_{\alpha=1}(N_\alpha(\vec{k})-\bar{N}_\alpha(\vec{k}))$$ which, in the non relativistic approximation (22) becomes $$\hat{q}^{nr}_0=e\int {{d^3\vec{k}}\over {(2\pi)^3}}\sum^2_{\alpha=1}(N_\alpha(\vec{k})-\bar{N}_\alpha(\vec{k})).$$ It is clear that, as its relativistic counterpart, when $\hat{q}_0^{nr}$ is applied to a state with $n$ particles (electrons) and $m$ antiparticles (positrons), the resulting eigenvalue is $(n-m)e$.

\

5. {\bf Fiber bundle description}

\

There is a vector space isomorphism between $M^4$, the Minkowski spacetime, and $H(2)$, the space of 2$\times$2 hermitian matrices: if $x\in M^4$ is any four-vector, then there is asociated the matrix given by $$\hat{x}=x_0I+\vec{x}\cdot \vec{\sigma}=x_\mu \sigma_\mu \eqno{(38a)}$$ where $I=\sigma_0$ is the unit 2$\times$ 2 matrix.

\

Conversely, given the 2$\times$2 matrix $\hat{x}$, $$x_\mu={{1}\over{2}}tr(\hat{x}\sigma_\mu). \eqno{(38b)}$$

\

As is well known, $SL_2(\C)$ is the universal covering group of $\cal{L}$$_0$, the connected component of the Lorentz group $^{9}$. On the other hand, the complexification of the full Lorentz group $\cal{L}$ is given by  $$({\cal L})^c=\{\Lambda\in \C(4)\vert \Lambda \eta \Lambda ^T=\eta\}, \eqno{(39)}$$ where $\eta=diag(1,-1,-1,-1)$ is the relativistic spacetime metric. Then a 2 $\to$ 1 covering group of the complexification of ${\cal L}_0$, $({\cal L}_0)^c$, is $SL_2(\C)\times SL_2(\C)$ $^{10}$.

\

Notice that if ${\cal L}$ is complexified, then the spacetime coordinates $x^\mu$ are not necessarily real but can take complex values; the same applies for any subgroup of ${\cal L}$. {\it This is a mathematical fact and no physical meaning is attributed here to the imaginary part of the coordinates.}

\

In a similar way, the complexifications of the groups $O(3)$ and its restriction $SO(3)$ are $$O(3)^c =\{A\in \C(3)\vert AA^\dagger=1\} \eqno{(40a)}$$ and $$SO(3)^c=\{A\in O(3)^c\vert det(A)=1\}. \eqno{(40b)}$$

\

We have the following inclusions of principal bundles:$$\matrix{\Z_2  &  \Z_2  &  (\Z_2)^2  & (\Z_2)^3\cr \downarrow  & \downarrow  & \downarrow & \downarrow \cr SU(2)\times SU(2) & \buildrel {\iota_c}\over \longrightarrow  S_{\pm}U(2)\times S_{\pm}U(2) \buildrel {\iota^\prime_c}\over \longrightarrow & (S_{\pm}U(2)\times \Z_2)^2 \buildrel {\iota^{\prime \prime}_c} \over \longrightarrow & (S_{\pm}U(2)\times \Z_2)^2\times \Z_4\cr \downarrow \pi_c & \downarrow \Pi_c & \downarrow q_c  & \downarrow Q\cr SO(3)^c  \buildrel {\bar{\iota}_c}\over \longrightarrow & O(3)^c  \buildrel {\bar{\iota}^\prime_c} \over \longrightarrow & O(3)^c\times \Z_2  \buildrel {\bar{\iota}^{\prime \prime}_c} \over \longrightarrow & O(3)^c\times (\Z_2)^2 \cr} \ . \eqno{(41)}$$ 

This is the complexification of the real case $^{6}$, where the last bundle in the right hand side contains the charge conjugation operation in the group $\Z_4\cong \{1,-1,\iota, -\iota\}$, $\iota^2=-1$, through the identification $$C_{nr}\equiv \iota. \eqno{(42)}$$ For $(A,B)\in SU(2)\times SU(2)$ the projection (group epimorphism) $\pi_c$ is given by $$\pi_c(A,B)(\hat{\vec{x}})=A\hat{\vec{x}}B^\dagger \eqno{(43)}$$ where $\hat{\vec{x}}=x_i\sigma_i$. Defining $$(\pi_c(A,B))^\dagger(\hat{\vec{x}})=A^\dagger \hat{\vec{x}}B \eqno{(44)}$$ one has $$\pi_c(A,B)\circ (\pi_c(A,B))^\dagger (\hat{\vec{x}})=\pi_c(A,B)((\pi_c(A,B))^\dagger (\hat{\vec{x}}))=\pi_c(A,B)(A^\dagger \hat{\vec{x}}B)=A(A^\dagger\hat{\vec{x}}B)B^\dagger=(AA^\dagger)\hat{\vec{x}}(BB^\dagger)$$ $=\hat{\vec{x}}$, in agreement with (40a) and (40b). 

$\Pi_c$ is as $\pi_c$ with $A,B\in S_{\pm}U(2)$. 

For $((A,\mu),(B,\nu))\in (S_{\pm}U(2)\times \Z_2)\times (S_{\pm}U(2)\times \Z_2)$, $$q_c((A,\mu),(B,\nu))=(\Pi_c(A,B),\mu\nu). \eqno{(45)}$$ It can be easily shown that $q_c$ is a group homomorphism.

Finally, the group homomorphism $Q$ is given by $$Q((A,\mu),(B,\nu),\lambda)=(q_c((A,\mu),(B,\nu)),\tilde{\eta}) \eqno{(46)}$$ where $$\tilde{\eta}=\{ \matrix{1 \ if \ \lambda=1 \ or \ -1 \cr -1 \ if \ \lambda=\iota \ or \ -\iota \cr} \ .$$ The existence of a short exact sequence of groups $$0\to G_1\buildrel {\alpha}\over\longrightarrow G_2 \buildrel {p}\over \longrightarrow G_3 \to 0$$ (where $ker(p)=Im(\alpha)\cong G_1$) $^{11}$ allows us to construct the principal bundle $G_1\to G_2\buildrel {p}\over \longrightarrow G_3$ where $G_1$ is the fiber, $G_2$ is the total space and $G_3$ is the base space. Clearly, only for $A=B=I\in \Z_2$, $\pi_c(A,B)(\hat{\vec{x}})=\hat{\vec{x}}$, and $\Pi_c(A,B)(\hat{\vec {x}})=\hat{\vec{x}}$; then $\Z_2$ is the fiber of the first two bundles.

From (45), the kernel of $q_c$ {\it i.e.} the fiber of the third bundle is given by $$ker(q_c)=\{((I,1),(I,1)),((I,-1),(I,-1)),((-I,1),(-I,1)), ((-I,-1),(-I,-1))\} \eqno{(47)}$$ with the following multiplication table: $$\matrix{& ((I,1),(I,1)) & ((I,-1),(I,-1)) & ((-I,1),(-I,1)) & ((-I,-1),(-I,-1)) \cr ((I,1),(I,1)) & ((I,1),(I,1)) & ((I,-1),(I,-1)) & ((-I,1),(-I,1)) & ((-I,-1),(-I,-1) \cr ((I,-1),(I,-1)) & ((I,-1),(I,-1)) & ((I,1),(I,1)) & ((-I,-1),(-I,-1)) & ((-I,1),(-I,1) \cr ((-I,1),(-I,1)) & ((-I,1),(-I,1)) & (-I,-1),(-I,-1)) & ((I,1),(I,1)) & ((I,-1),(I,-1)) \cr ((-I,-1)(-I,-1) & ((-I,-1),(-I,-1) & ((-I,1), (-I,1)) & ((I,-1), (I,-1)) & ((I,1),(I,1)) \cr}$$ With the isomorphism $$((I,1),(I,1))\to (1,1), \ ((I,-1),(I,-1))\to ((1,-1), \ ((-I,1,-I,1))\to (-1,1),$$ $$ ((-I,-1),(-I,-1))\to (-1,-1),$$ $ker(q_c)\cong\Z_2\times \Z_2$, the Klein group.

In a similar way, $$ker(Q)=\{((I,1),(I,1),1),((I,-1),(I,-1),1), ((-I,1),(-I,1),1), ((-I,-1),(-I,-1),1),$$ $$ ((I,1),(I,1),-1),((I,-1),(I,-1),-1), ((-I,1),(-I,1),-1), ((-I,-1),(-I,-1),-1)\} \eqno{(48)}$$ with multiplication table $$\matrix{& ((I,-1),(I,-1),1) & ((-I,1),(-I,1),1) & ((-I,-1),(-I,-1),1) \cr 
((I,-1),(I,-1),1) & ((I,1),(I,1),1) & ((-I,-1),(-I,-1),1) & ((-I,1),(-I,1),1) \cr ((-I,1),(-I,1),1) & ((-I,-1),(-I,-1),1) & ((I,1),(I,1),1) & ((I,-1),(I,-1),1) \cr ((-I,-1),(-I,-1),1) & ((-I,1),(-I,1),1) & ((I,-1),(I,-1),1) & ((I,1),(I,1),1) \cr }$$ where for simplicity the row and column corresponding to the identity $((I,1),(I,1),1)$ have been supressed and also those corresponding to -1 in $\Z_4$. The isomorphism $$((I,1),(I,1),1)\to (1,1,1), \ ((I,-1),(I,-1),1)\to (1,-1,1),$$ $$ ((-I,1),(-I,1),1)\to (-1,1,1), \ ((-I,-1),(-I,-1),1)\to (-1,-1,1),$$ $$((I,1),(I,1),-1)\to (1,1,-1), \ ((I,-1),(I,-1),-1)\to (1,-1,-1),$$ $$ ((-I,1),(-I,1),-1) \to (-1,1,-1), \ ((-I,-1),(-I,-1),-1) \to (-1,-1,-1)$$ makes $ker(Q)\cong (\Z_2)^3$.

\

The identification of $\iota$ with $C_{nr}$, the definition of $Q$, and the complexification of the space coordinates, allows us to define $$Q((I,1),(I,1),\pm 1)(\hat{\vec{x}})=\hat{\vec{x}} \eqno{(49a)}$$ and $$Q((I,1),(I,1),\pm \iota)(\hat{\vec{x}})=(\hat{\vec{x}})^* \eqno{(49b)}$$ where $(\hat{\vec{x}})^*=(x_i)^*\sigma_i$. So, the complexification of spacetime, in the context of a fiber bundle description, allows for a direct action of the charge conjugation matrix (20a) on the space coordinates, namely of $C_{nr}$ on $\hat{\vec{x}}\in \C^3$.

\

{\bf Acknowledgement}

\

M. S. was partially supported by the project PAPIIT IN103505, DGAPA-UNAM, M\'exico. Also, M. S. thanks the Abdus Salam International Centre of Theoretical Physics, Trieste, for support through the Net 35, and the hospitality of the ICIMAF, La Habana, Cuba, where part of this work was performed.

\

{\bf References}

\

1. V. B. Berestetskii, E. M. Lifshitz, and L. P. Pitaevskii, {\it Quantum Electrodynamics, Landau and Lifshitz Course of Theoretical Physics, Vol. 4}, 2nd. edition, Pergamon Press, Oxford (1982): p. 45.

\

2. E. S. Abers, {\it Quantum Mechanics}, Pearson Education, New Jersey (2004): p. 102.

\

3. I. I. Bigi, and A. I. Sanda, {\it CP Violation}, Cambridge University Press, Cambridge (2000): pp. 14, 18.

\

4. J. D. Bjorken, and S. D. Drell, {\it Relativistic Quantum Mechanics}, Mc Graw-Hill, New York (1964): p. 11.

\

5. M. Socolovsky, The CPT Group of the Dirac Field, {\it International Journal of Theoretical Physics}  {\bf 43}, 1941-1967 (2004); arXiv: math-ph/0404038.

\

6. D. B. Cervantes, S. L. Quiroga, L. J. Perissinotti, and M. Socolovsky, Bundle Theory of Improper Spin Transformations, {\it International Journal of Theoretical Physics} {\bf 44}, 267-276 (2005); arXiv: quant-ph/0410079.

\

7. J. A. de Azc\'arraga and J. M. Izquierdo, {\it Lie groups, Lie algebras, cohomology and some applications in physics}, Cambridge University Press, Cambridge (1995): pp. 154-155.

\

8. L. Wolfenstein, and D. G. Ravenhall, Some Consequences of Invariance under Charge Conjugation, {\it Physical Review} {\bf 88}, 279-282 (1952).

\

9. S. Sternberg, {\it Group Theory and Physics}, Cambridge University Press, Cambridge (1994): pp. 2, 6.

\

10. R. F. Streater, and A. S. Wightman, {\it PCT, Spin and Statistics, and All That}, Benjamin, New York (1964): p. 14.

\

11. S. Mac Lane, and G. Birkoff, {\it Algebra}, 2nd. ed., Macmillan Pub. Co., New York (1979): p. 413.

\

\

\

\

\

e-mails: cabo@icmf.inf.cu, daliac@nucleares.unam.mx, hugo@icmf.inf.cu, socolovs@nucleares.unam.mx

\end